\documentclass[a4paper]{jpconf}
\usepackage{graphicx}

\def\C60{C$_{60}$}
\def\Csi{Cs$^+$}
\def\Cs3C60{Cs$_3$C$_{60}$}
\def\cm-1{cm$^{-1}$}
\def\T1u{$T_{1u}$}
\def\D2h{$D_{2h}$}
\def\Ci{C$_{60}^{3-}$}
\def\t1u{$t_{1u}$}

\begin{document}

\title{Mott localization in the correlated superconductor Cs$_3$C$_{60}$ resulting from the molecular Jahn-Teller effect}

\author{Katalin Kamar\'{a}s$^1$, Gy\"{o}ngyi Klupp$^1$, P\'{e}ter Matus$^1$, Alexey Y. Ganin$^2$, Alec McLennan$^2$, Matthew J. Rosseinsky$^2$, Yasuhiro Takabayashi$^3$, Martin T. McDonald$^3$, Kosmas Prassides$^3$}
\address{$^1$ Institute for Solid State
Physics and Optics, Wigner Research Centre for Physics, Hungarian
Academy of Sciences, P.O. Box 49, H-1525 Budapest, Hungary}
\address{$^2$ Department of Chemistry, University of Liverpool,
Liverpool L69 7ZD, UK}
\address{$^3$ Department of Chemistry, Durham University, Durham DH1 3LE, UK}
\ead{kamaras.katalin@wigner.mta.hu}

\begin{abstract}

\Cs3C60 is a correlated superconductor under pressure, but an insulator under ambient
conditions. The mechanism causing this insulating behavior is the combination of Mott localization and the dynamic Jahn-Teller effect. We show evidence from infrared spectroscopy for the dynamic Jahn-Teller distortion.
The continuous change with temperature of the splitting of infrared lines is typical Jahn-Teller behavior, reflecting the change in population of solid-state conformers. We conclude that the electronic and magnetic solid-state properties of the insulating state are controlled by molecular phenomena. We estimate the time scale of the dynamic Jahn-Teller effect to be above 10$^{-11}$ s and the energy difference between the conformers less than 20 cm$^{-1}$.

\end{abstract}

\section{Introduction} Trivalent fulleride salts belong to the family of correlated
superconductors, similar to high temperature superconducting cuprates
\cite{takabayashi09,tosatti09}. They show a dome-shaped $T_c$ (superconducting transition temperature) versus lattice
constant phase diagram \cite{ganin08} with an antiferromagnetic insulating state above a critical lattice constant \cite{takabayashi09}. The transition from metallic/superconducting to insulating behavior is generally attributed to Mott localization. Fullerides showing similar behavior have been known for a long time.
Metallic fcc (face centered cubic) mixed alkali
Cs$_x$A$_{(3-x)}$\C60 (A=K,Rb) phases have been prepared, which were in the lattice
constant range where $T_c$ decreases upon expansion \cite{dahlke00}.
Guest molecules, like cubane \cite{pekker05} or ammonia
\cite{rosseinsky93} were also used to expand the lattice. This approach leads in most cases to a change in crystal structure, inducing a crystal field with lower symmetry which distorts the fulleride molecular ions. Metal-insulator transitions in these compounds have been confirmed \cite{allen96, iwasa96}, but since these occur in a low-symmetry crystal field, the external and internal contributions could not be unambiguously distinguished. \Cs3C60 is in this respect a key compound. It is an expanded structure, shows a metal-to-insulator transition upon application of pressure \cite{ganin08,ihara10} and retains its cubic structure in the whole pressure range.

\Cs3C60 has two cubic polymorphs, an ordered A15 \cite{ganin08} and a
merohedrally disordered fcc one \cite{ganin10}. At ambient pressure they are insulators and magnetic with S=1/2, but turn into metals upon application of pressure \cite{ihara11}, exhibiting superconducting transition temperatures T$_c$ up to 38 K \cite{ganin08}. The metal-insulator transition upon increasing distance between the fulleride ions and the localized magnetic moments point to strong electron-electron correlation and consequently, Mott localization.

The localization mechanism was suggested to involve a Jahn-Teller distortion \cite{capone09}, which is difficult to prove experimentally. Atomic displacements in the distorted molecular ions have been estimated to fall in the range of 0.05 {\aa}ngstroms, hard to detect even by sensitive neutron scattering experiments. However, since the distortion involves a symmetry change, vibrational spectroscopy can be a sensitive indicator. The T$_{1u}$(4) infrared-active mode at 1429 cm$^{-1}$ is most widely used to detect both charge and symmetry changes in C$_{60}$ \cite{kamaras97b}. From symmetry considerations for an I$_h$ point group object placed in a cubic environment, it follows that this mode will not show splitting caused by the environment. Therefore, all splittings found can be directly related to the symmetry reduction following from the molecular Jahn-Teller effect.

\section{The Jahn--Teller effect in solid fullerides}

\begin{figure}[h]
\includegraphics[width=23pc,trim=0 20 0 10]{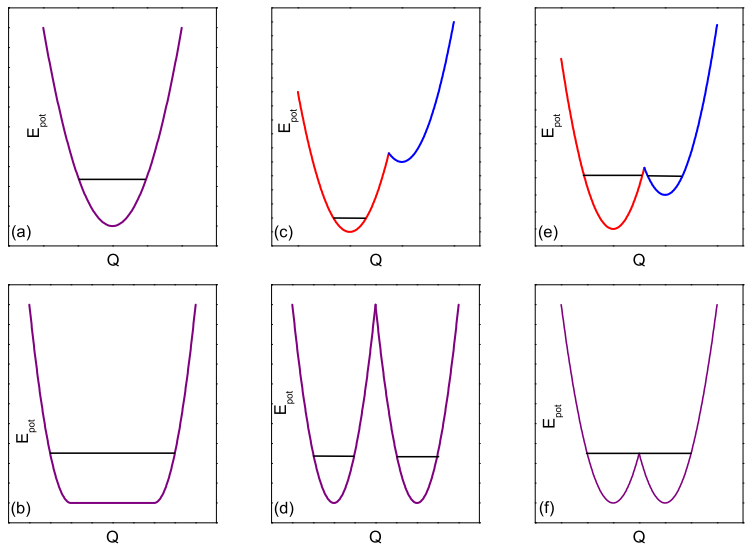}
\caption{\label{fig:Epot}Possible potential energy surfaces of the
\Ci\ molecular ion in \Cs3C60. (a) The possible D$_{2h}$ distortions in
all directions have the same energy. (b) No energy barrier between
differently directed distortions causes an average I$_h$ symmetry.
(c, e, f) Two inequivalent directions of differing energy separated by a
small energy barrier. (c) Static distortion in one of
the directions. (d) Two inequivalent directions of similar energy
separated by a large energy barrier. A small
energy difference between the two wells is possible. Static distortion. (e)
Dynamic distortion with two solid-state conformers with differing
population. (f) Hindered pseudorotation between differently directed
distortions.}
\end{figure}

According to theoretical calculations \cite{auerbach94} the symmetry
of the Jahn--Teller distortion of \Ci\ is D$_{2h}$. The potential
energy surface of a single distortion is schematically illustrated in
Fig.~\ref{fig:Epot}~(a). The possible combinations of fifteen equivalent twofold axes among the symmetry elements of icosahedral \C60\ define thirty equivalent distortions among which the molecule can move.
The corresponding motion is called pseudorotation (see the movie in the Supplementary
Information). The molecule performing pseudorotation is dynamically
distorted. If there is no barrier between the different D$_{2h}$
distortions (free pseudorotation), the molecular
symmetry will be restored to I$_h$ in time average
\cite{bersuker06} (see Fig.~\ref{fig:Epot}\ (b)). Introducing a
barrier between the D$_{2h}$ distortions results in
hindered pseudorotation \cite{bersuker06} (see
Fig.~\ref{fig:Epot}~(e) and (f)). If the time scale of a
spectroscopic measurement is shorter than that of pseudorotation, the
D$_{2h}$ distortion can be detected. The time scale of infrared excitations being of the order of 10$^{-11}$ s,
spectra corresponding to distorted states correspond to pseudorotations slower than this value. The last possibility is that
with a barrier high enough to cause static distortion along one special axis
\cite{bersuker06} (see Fig.~\ref{fig:Epot}~(c) and (d)).

In the solid one has to take into account the effect of the crystal field determined by the potential created by the neighboring \Csi\ ions.
The undistorted site symmetry is T$_h$ in these cubic crystals, so the crystal field alone would not lower the symmetry of the fulleride ions below that point group.\cite{klupp12}. The number of twofold axes along which Jahn-Teller distortion can occur is reduced to three, and if one considers the \Ci\ ion in a T$_h$ point group there will be just two inequivalent directions. This is why two kinds of distortions appear in Fig.~\ref{fig:Epot}~(c)-(f).

Thermal expansion results in weaker crystal field
acting towards equalizing the two kinds
of distortions. In addition, at higher temperature the higher energy states become increasingly populated. Thus, for example, heating will lead from the state in Fig.~\ref{fig:Epot}~(c) to (e), and further to (f).
Raising the temperature will also lower the barriers between potential energy wells (best seen in
Fig.~\ref{fig:Epot}~(d)) and they can be more easily overcome.

\section{Experimental}

The \Cs3C60 samples used in this study were prepared by solution chemistry routes as described elsewhere \cite{ganin08,ganin10}. Sample composition and structure were determined by x-ray diffraction. Infrared measurements were done on both fcc-rich and A15-rich materials, in KBr pellets with 0.25 cm$^{-1}$ resolution \cite{klupp12}.

\section{Signatures of the Mott insulating state}

\begin{figure}[h]
\includegraphics[width=23pc,trim=0 20 0 0]{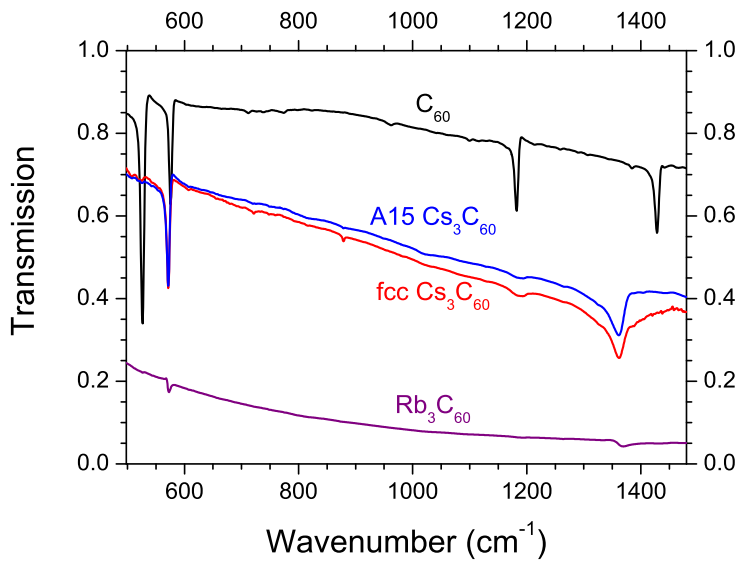}\hspace{2pc}%
\begin{minipage}[b]{12pc}\caption{\label{fig:Rb}Infrared spectrum of A15 and fcc \Cs3C60 in the range of
intramolecular vibrations at room temperature. The room temperature
spectra of insulating \C60 and metallic Rb$_3$C$_{60}$ are included for comparison. The metallic character influences both the background and the lineshape.}
\end{minipage}
\end{figure}

The measured infrared spectra of both polymorphs together with that of \C60 and of metallic Rb$_3$C$_{60}$ are shown in
Fig.~\ref{fig:Rb}. The spectrum of pure \C60
consists of four lines, a consequence of the high symmetry of the fullerene ball \cite{kratschmer90}. In accordance with previous data \cite{pichler94}, the highest
frequency mode softens due to the charge added on the fullerene, showing approximately the same shift in \Cs3C60 and in Rb$_3$C$_{60}$. The lineshape of this mode, however, is different for Rb$_3$C$_{60}$ and for \Cs3C60. Rb$_3$C$_{60}$ has a Fano lineshape \cite{iwasa93} as a
consequence of coupling between vibrational states and the electronic continuum of the same energy. The non-Fano lineshape in both polymorphs of \Cs3C60 signals the absence of metallic electrons, therefore these materials are insulators,
in agreement with previous measurements \cite{takabayashi09}.
Metallic electrons also cause a continuous background
absorption, which is indeed present in Rb$_3$C$_{60}$, but not in either of the two \Cs3C60 polymorphs.

Another signature of the insulating state is that it facilitates the appearance of the Jahn--Teller effect: molecular ions can suffer a Jahn--Teller distortion only if their electrons are localized. The asymmetrically broadened lineshape in both \Cs3C60 compounds (Fig.~\ref{fig:Rb}) indicates splitting of the vibrational lines, a consequence of the symmetry reduction owing to the Jahn--Teller distortion \cite{klupp12}. To further study the temperature and the polymorph dependence of the splitting, temperature dependent measurement were performed on both \Cs3C60 polymorphs.

\section{Jahn--Teller effect in \Cs3C60}

\begin{figure}[h]
\includegraphics[width=23pc,trim=0 0 0 0]{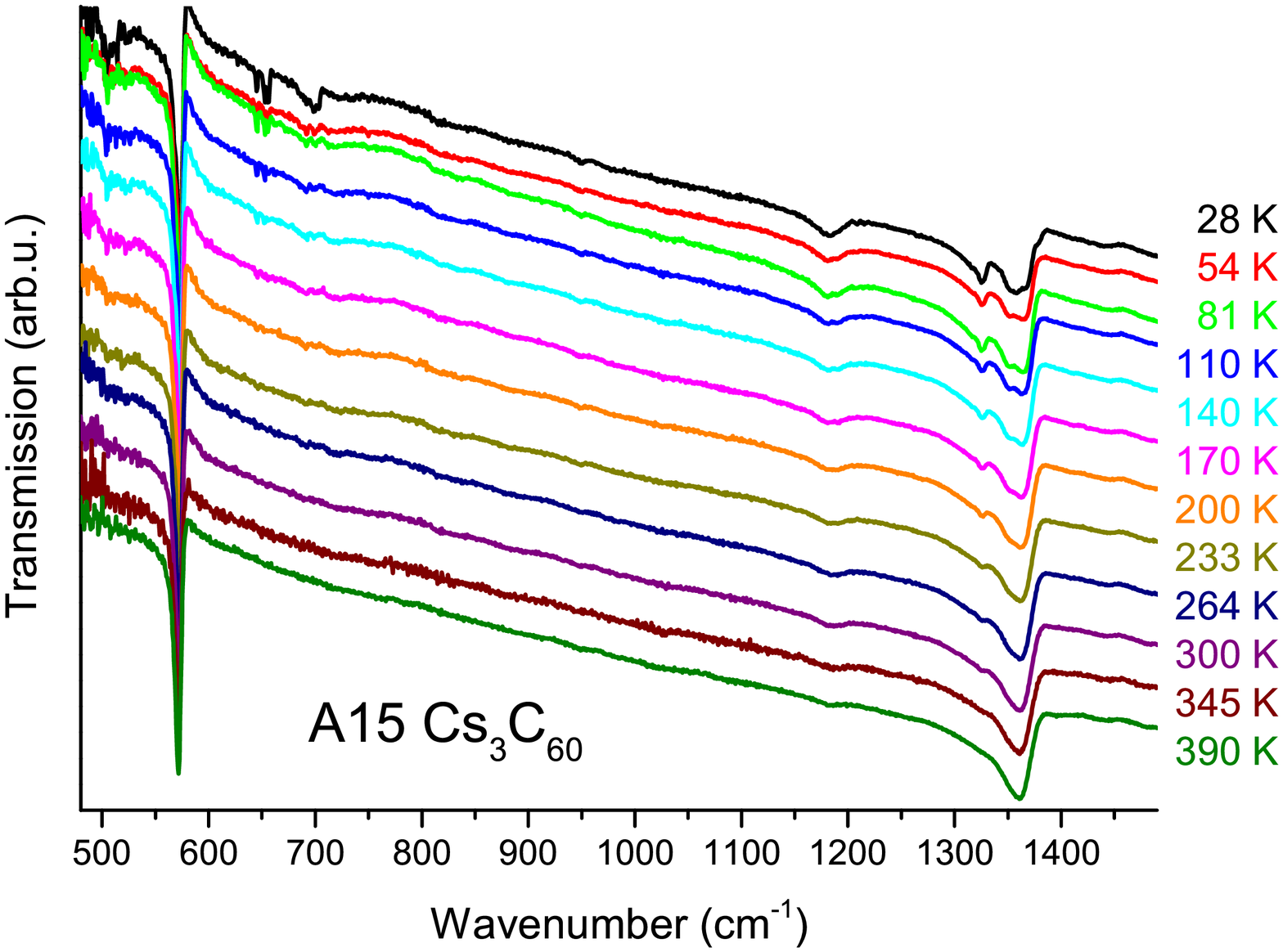}\hspace{2pc}%
\begin{minipage}[b]{12pc}
\includegraphics[width=9pc,trim=0 0 0 0]{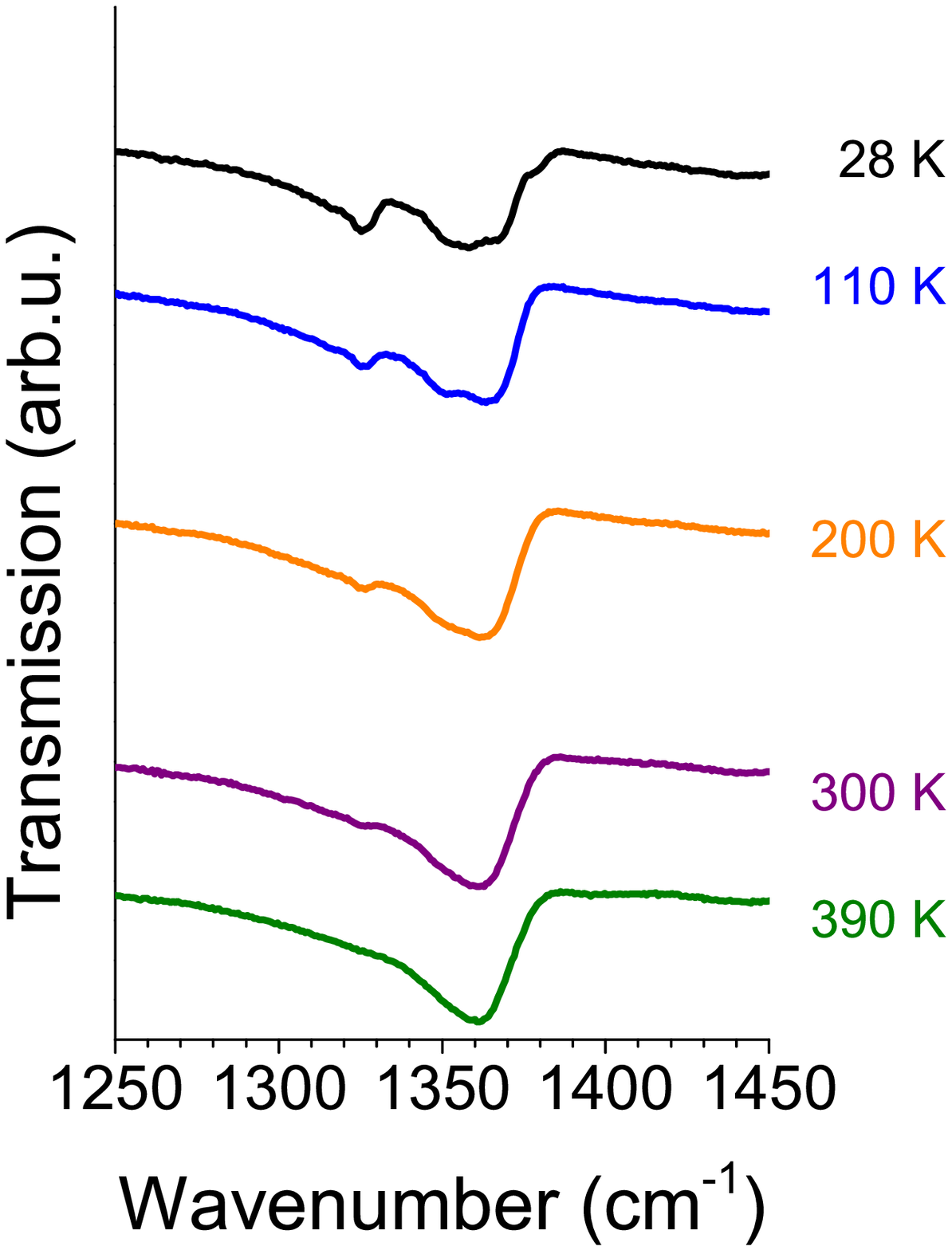}
\caption{\label{fig:A15}Temperature-dependent IR spectra of A15 \Cs3C60 (left: full spectrum, right: range of T$_{1u}$(4) and G$_u$(6) modes). The splitting changes gradually.}
\end{minipage}
\end{figure}

\begin{figure}[h]
\includegraphics[width=23pc,trim=0 0 0 0]{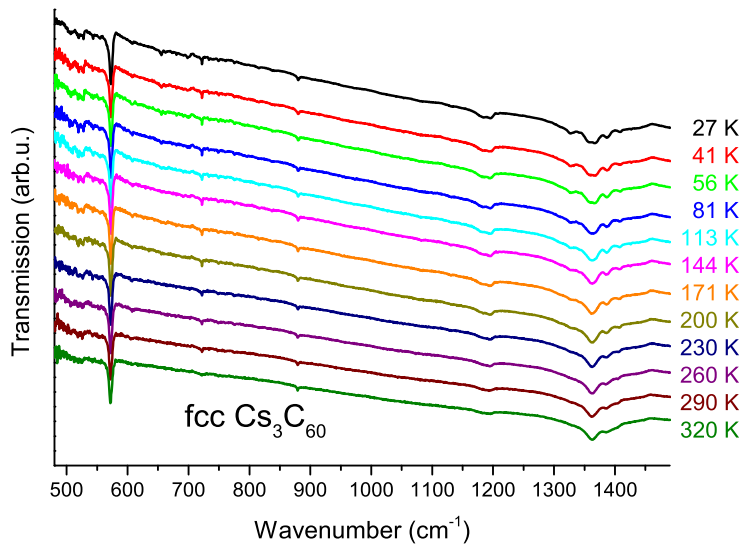}\hspace{2pc}%
\begin{minipage}[b]{12pc}
\includegraphics[width=9pc,trim=0 0 0 0]{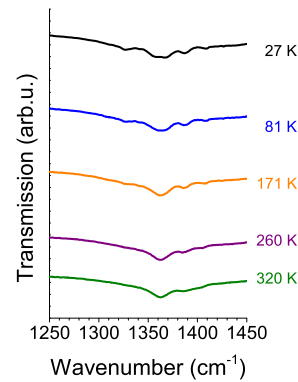}
\caption{\label{fig:fcc}Temperature dependent IR spectra of fcc \Cs3C60. The splitting changes gradually.}
\end{minipage}
\end{figure}

Figures~\ref{fig:A15} and \ref{fig:fcc} illustrate the
gradual changes in the splitting, characteristic of both polymorphs (although not exactly identical). Newly activated low intensity modes between 600-800~\cm-1 can also be observed at low temperature. These changes indicate that the symmetry lowering is different in the two \Cs3C60 polymorphs. The weak new modes disappear at differing temperatures underlining the smooth nature of the transition. As we will see, this is a sign of temperature-dependent solid-state conformers, implying a dynamical Jahn--Teller effect \cite{bersuker06}.

The low-temperature multiple peak structure and its dependence on crystal structure is a result of the interplay between crystal field and Jahn--Teller distortion. The \Csi\ ions that surround the \Ci\ ion constitute a cubic environment and do not cause splitting in the T$_{1u}$ modes. However, the energies of molecular distortions, that are equivalent in a free molecular ion, can differ slightly in a crystal when pointed at different crystallographic directions. The observation of the signatures of the Jahn--Teller distortion in the vibrational spectra rules out the case of Fig.~\ref{fig:Epot}~(b) (free pseudorotation). Statically ordered distortions in the lattice with the same energy (Fig.~\ref{fig:Epot}~(a)) would lead to a
temperature-independent spectrum consisting of one set of lines (maximum 3+4, for triply degenerate T$_{1u}$ and quadruply degenerate G$_u$, respectively) and is also contrary to observation. The number of possible allowed IR lines (in principle 1* or 2*(3+4)) corresponds to the number of populated potential energy wells (1 in Fig.~\ref{fig:Epot}~(a-c) and 2 in Fig.~\ref{fig:Epot}~(d-f)). According to the experiments, this number continously decreases on heating. This finding can be qualitatively explained by a model where the population of the wells changes with temperature.

The simplest of these models is that of the static-to-dynamic transition (from Fig.~\ref{fig:Epot}~(d) to (f)) on heating. At low temperature the differently directed distortions would be frozen in with static and disordered distortion. Thermal expansion lowers the barrier and the increased thermal energy from pseudorotation helps to overcome it. In this case the transition would be located in a considerably narrower temperature range, but we observe a continuous change over the whole temperature range we measured. The model based on Fig.~\ref{fig:Epot}~(d), where the barrier is too high for a transition even at high temperature, can also be rejected based on the measured temperature-dependent infrared data. A  transition could also happen from the static distortion shown in Fig.~\ref{fig:Epot}~(c) to the dynamic ones Fig.~\ref{fig:Epot}~(e) and (f). The number of lines should first increase on heating, then start to decrease. As we cannot prove such a behaviour, this case can also be ruled out.

This leaves us with the dynamic distortions Fig.~\ref{fig:Epot}~(e) and (f), the latter developing from the former on increasing temperature. This situation corresponds to different solid-state conformers with different occupational probability \cite{klupp12}. On heating, both their energy difference and their populational difference gradually disappears \cite{bersuker06}, as observed in the spectra. Fewer kinds of distortions mean fewer lines in the infrared spectra.
The two equilibrium conformations with a low barrier (Fig. \ref{fig:Epot}e) are connected by hindered pseudorotation \cite{bersuker06}. The fact that we observe the distorted states puts the lower limit of the time scale of this pseudorotation to 10$^{-11}$ s, and the lowest temperature being 28 K can be reconciled with an upper limit of 20 cm$^{-1}$ for the barrier between the two conformers.

\section{Conclusions}

We studied the temperature-dependent infrared spectra of both polymorphs of the correlated superconductor \Cs3C60\ in their normal state. The vibrational signatures reveal the insulating state of these materials, and establish the interplay of Mott localization and the Jahn-Teller effect in forming this state. In addition, the crystal field influences the vibrational levels making the spectra temperature- and polymorph-dependent. The presence of temperature-dependent solid-state conformers validates the proof of the dynamic Jahn--Teller effect.

\ack
Funding for this research was provided by the Hungarian
National Research Fund (OTKA) (Grant No. 75813), the Engineering and
Physical Sciences Research Council(Grant No. EP/G037132 and
EP/G037949), and the EU-Japan project LEMSUPER (Grant No.
NMP3-SL-2011-283214).

\noindent\textbf{Supplementary Information} accompanies this paper.

\section*{References}

\bibliography{Cs3ir_jt}
\bibliographystyle{iopart-num}

\end{document}